# Backscattering of MeV Electrons. An Analysis of Tabata's Experiment with Geant4

Sylvian Kahane

*Abstract* — A series of simulations were conducted with Geant4 in order to verify the electron backscattering experiments performed by Tabata in the low Z elements of Be, C and Al. In general, a quite good agreement was obtained by carefully choosing the physics lists employed. These results invalidate the claim made before by Kirihara *et al* about the presence of experimental errors in Tabata's work.

*Keywords* — backscattering coefficient, electron scattering, EGS5 simulation, medium (MeV) energy.

I. INTRODUCTION

The field of the electron backscattering is a very active one, rich both in experiments and analysis works. Kim *et al.* [1], for example, collected data from some 3000 experiments, covering an energy range from 78 eV to 22.2 MeV. Knowledge about backscattering of electrons is important in many fields. For example, in medical physics, mainly in the kilovolt range of energies, it is of critical importance in establishing accurate irradiation doses to the patients. The papers of Ali and Rogers [2] and a previous work by the present author [3], are dealing with this range of energy. Other fields of interest are surface science, electron field microscopy and many others. The experimental work is difficult and prone to errors.

In the energy region of a couple of MeV, Tabata [4] published already 50 years ago a comprehensive work which stands out by its quality. Angular distributions were measured for a full range of backward angles, at 5 energies between 3.24 to 14.1 MeV, from a number of targets which included heavy, medium and light nuclei. More recently, Kirihara *et al.* [5] used the Monte Carlo code EGS5 to try to reproduce the experimental results of Tabata and at a same time to validate the electron physics included in the code. While for the heavy nuclei the agreement was fair to good, for the light one's substantial discrepancies where observed. Especially in beryllium, a factor higher that 2 was obtained for EGS5/Tabata ratio. This discrepancy prompted the above authors in proposing the existence of an experimental problem in the calibration method used in the work of Tabata.

The backscattering coefficient is defined as the number of backscattered electrons over the number of total electrons impinging on the target. Experimentally, one does not count electrons but measures currents. Tabata used an Ionization Chamber which amplifies the number of electrons and, hence, his measured current needed a calibration factor. It was obtained from the work of Wright and Trump [6], in which a Faraday cup, which does not disturb the number of electrons, was used.

The EGS Monte Carlo code has a long history beginning in the sixties at SLAC and culminating with EGS4 [6] in the eighties. It is a very successful code with a good record in reproducing the experimental data. Today, it is followed by EGSnrc [8] at the NRC in Canada, which is keeping with the heritage of developing codes in a scripting language and subsequently translating to FORTRAN via an interpreter named MORTRAN (also developed at SLAC a long time ago, and aware only of the FORTRAN77 standard). Another group, based at KEK in Japan, developed some 10 years ago a fully FORTRAN version known as EGS5 [9]. As such it is more appropriate for introducing features available only in the more modern FORTRAN standards.

The solution proposed by Kirihara *et al.* [1], namely the existence of an experimental problem in Tabata's work, contradicts the findings of an older paper of Benedito, Fernandez-Varea and Salvat [10] which presents a perfect agreement for both Aluminium and Gold with Tabata's results. Their work is computationally intensive, requesting solutions of the electron Dirac equation, with the ELSEPA [11] program, in an angular momentum expansion, reaching values of l up to 4000.

The present work aims at investigating the above contradiction by using the Geant4 Monte Carlo simulation toolkit [12]. Ref. [1] pointed out that the electron transport algorithms are using approximations

---

for the multiple-scattering, not necessarily working for all the energies and all the scattering angles. In general, the sampling of individual Coulomb collisions is avoided and, instead, the electron tracks are divided into segments [13]. The angular deflections and energy losses in successive segments are sampled from appropriate multiple-scattering distributions. The default distribution is based on the theory of Molière [13] but this of Goudsmit and Saunderson [15] is also available. In the EGS programs this approach is implemented in the PRESTA II [16] algorithm. Geant4 is more complex, consisting of many different models for different types of interactions. Users have to employ *physics lists* which aggregates the models to be used in a specific simulation. This approach creates serious problems for causal users, who do not possess in depth knowledge of the available physics *and* the C++ class structures to be employed for specific models. Fortunately, there are a fair number of ready-for-use physics list available for such users. More recently an effort was made in improving the multiple-scattering models [17].

## II. METHODS AND RESULTS

### A. The Geant4 Program

The program *TestEM5*, found in the *examples/extended/electromagnetic* folder of the 10.06.p02 Geant4 distribution, was used for calculation of the backscattering coefficient $\eta(E)$ and the angular distributions $d\eta(E)/d\Omega$. This program is devised to study the transmission, absorption and reflection of particles through a single, thin or thick, layer of material. It presents at output the backscattering coefficient in percent, and, using the preprogramed histogram #32 (for reflected and charged particles), the angular distribution of the coefficient, at a given electron incident energy. There is a problem with this histogram, if a user is asking for an x-axis in degrees a unit of $\pi/180=0.0174$ is set automatically, otherwise, when dealing only in radians, the unit is 1. When filling the corresponding histogram each score receives a weight of $w=1/\Delta\Omega$, programmed as:

$$\theta = \text{acos}(something)$$
$$w = (unit * unit)/(2\pi \sin(\theta)\, d\theta).$$

$\theta$ is in radians, from the acos statement, and does not need any unit. If the unit is defined as 1 there is no problem, but when the value is 0.0174 then $w$ is wrong. Furthermore $d\theta$ is the bin width of the histogram, usually defined in degrees. To transform it to radians one has to *multiply* by the unit, not to divide. Hence, the weight was calculated as $w = 1/(2\pi \sin(\theta)\, d\theta * unit)$ and not as given above. Finally, there are a number of similar histograms in the program, #12, #22, #32 and #42. Only #12 is actually normalized (divided by the total number of events) at the end of the run. A normalization was added also for #32.

The cuts were 10 and 100 μm for photons and electrons respectively, corresponding to 1 keV and 120 keV, depending on the material. This work concentrates on the low Z elements Be, C and Al, in which discrepancies with EGS5 calculations were found. Additionally, results for Gold are brought for comparison. The following physics lists were tried in order to obtain the best agreement:

TABLE I: GEANT4 PHYSICS LISTS USED IN THE CALCULATIONS

| Full name | Label | Details |
|---|---|---|
| emstandard_opt4 | opt4 | Standard electromagnetic interactions of Geant4 calculated with the best accuracy. The default MSC theory is that of Urban [18]. |
| emstandardGS | GS | Standard electromagnetic + Goudsmit & Saunderson[Error! Bookmark not defined.] theory for MSC. |
| emstandardWVI | WVI | Standard electromagnetic + Wetzel [19] theory of MSC. More demanding on the computing time. |
| empenelope | penelope | Physics from the Penelope [20] program, 2008 version. |
| emstandardSS | SS | Standard electromagnetic + Single scattering MSC. Gives excellent results according to Ref. **Error! Bookmark not defined.**. Abandoned due to excessive demands on my computational capabilities. |
| local | local | Private physics list included in the TestEM5 program. Standard EM physics with current 'best' options settings (according to the programmers). |

Not a single physics list (PL) was used for a given Z, rather the one producing a best result for a given $(E,\theta)$. One the other hand no cherry-picked choices were done, one PL for this angle and another for the next adjacent angle, rather a PL appropriate for a region of energies and angles was sought. The rationale is that no PL and no MSC formalisms are working universally equally well but rather for a region. The runs included $2\times10^6$ shots apart of the WVI case with only $1\times10^6$ shots. EGS5 is in general much more rapid then Geant4.

## B. Beryllium Results

For Beryllium the full data is shown, both as a function of energy and as angular distributions. Beryllium was measured at only 3 energies 3.24, 4.09 and 6.08 MeV. In Fig. 1 there are 7 frames, depending on the scattering angle, which show dη(E)/dΩ in units of $10^{-3}$ sr, and one frame showing the total backscattering coefficient η(E) in percent. The full lines are Geant4 calculation with WVI giving quite good agreement, the dashed lines are EGS5 calculations shown for comparison. The angular distributions are presented in Fig. 2 and a goodness of fit (GOF) test was performed. For this, a $\chi^2$ statistic (in every day parlance the chi square) was calculated from the experimental data, the errors and the calculated values It was compared to the critical value of a $\chi^2$ distribution with 6 degrees of freedom (number of measurements – 1) at the customary statistical significance of 5% which is 12.6. If the statistic is less than the critical value the null hypothesis is accepted otherwise rejected. The null hypothesis assumes that the measured data is fairly described by the calculations. Using what is called a *pvalue* is more transparent. The pvalue is the probability with which the calculated statistic appears in the above $\chi^2$ distribution. Given that in general the GOF test is done with a statistical significance of 5% it is enough to see if the pvalue is higher or lower than 5%. Of course, for a higher pvalue the agreement is better than for a smaller one, but the test itself is on/off – accepted/rejected. Fig. 2 presents also the pvalues, in all the 3 cases the null hypothesis is accepted even if at 6.08 MeV somehow barely, but it can be inferred from the top of the figure that still is much better than EGS5.

## C. Carbon Results

Carbon was measured at all the 5 energies and only the angular distributions are presented in 5 panels and the total η(E) in panel #6. The opt4 and the local PLs are doing quite a good job here apart for the case at 10.1 MeV where the fit is bad, as can be seen with the naked eye without the need of a p-value. EGS5 gives a bad fit as already was found by Kirihara.

From the above examples one can get the wrong impression that Geant4 will somehow solve all the problems. This impression is partially wrong, there are PLs who are not able to reproduce the experimental data. An example is given in Fig. 4. The Goudsmit-Saunderson (GS) is performing remarkably bad (as bad as EGS5) in reproducing the backscattering coefficient. On the other hand, opt0 and penelope are performing apparently remarkably good. But one should use all the available data in order to judge. Looking at 10.1 MeV angular distribution, the one who was rejected in Fig. 3, both are rejected, opt0 with a pvalue much lower but penelope with a higher pvalue. Not shown is the rejection suffered by penelope at 6.08 MeV with a p-value of 0.15%. Hence, all the PLs of this figure are performing pretty bad, Geant4 is not a superman compared with EGS5.

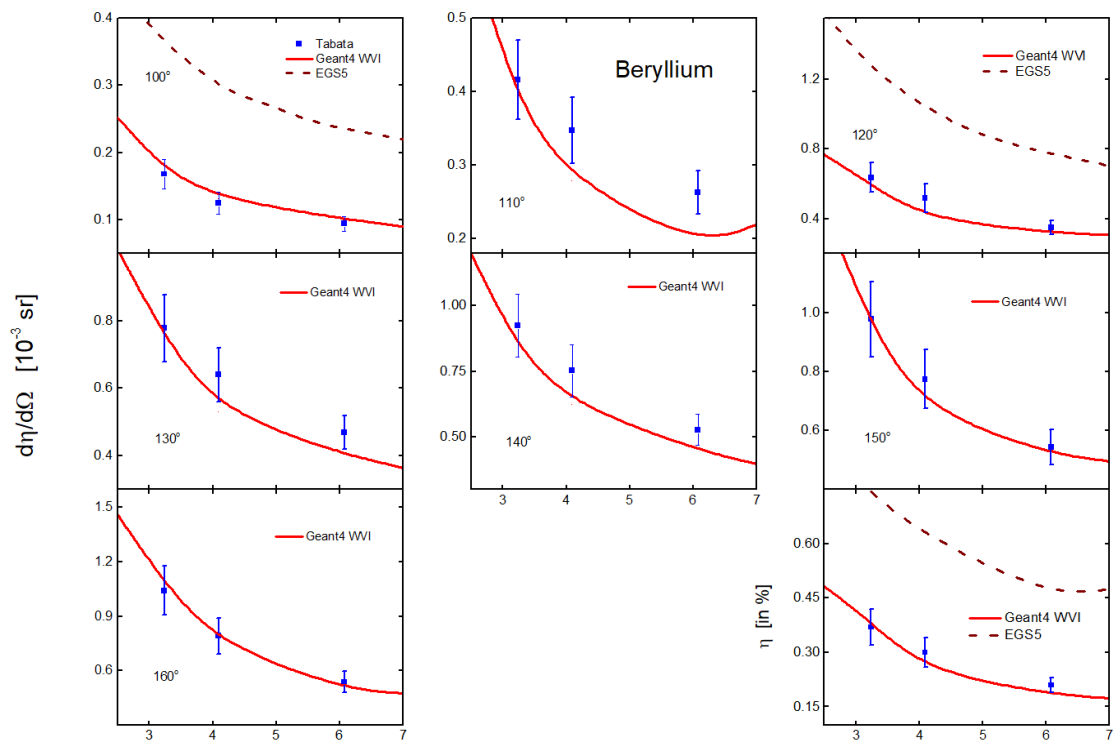

Fig. 1. Beryllium data as a function of the incident electron energy.

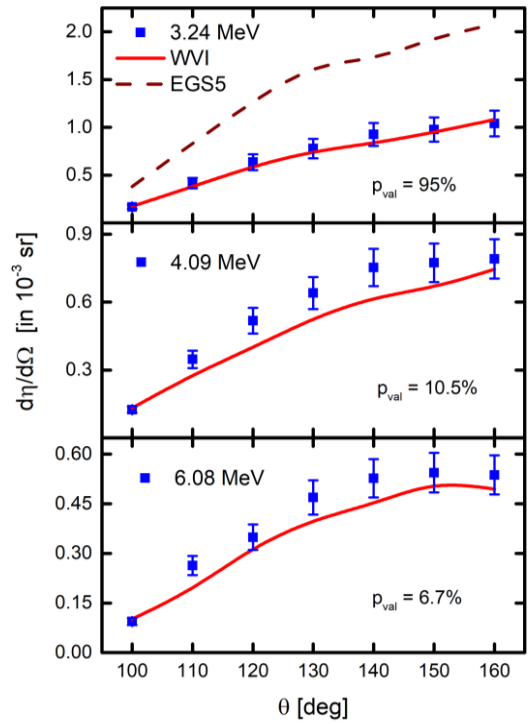

Fig. 2. Angular distributions in Beryllium.

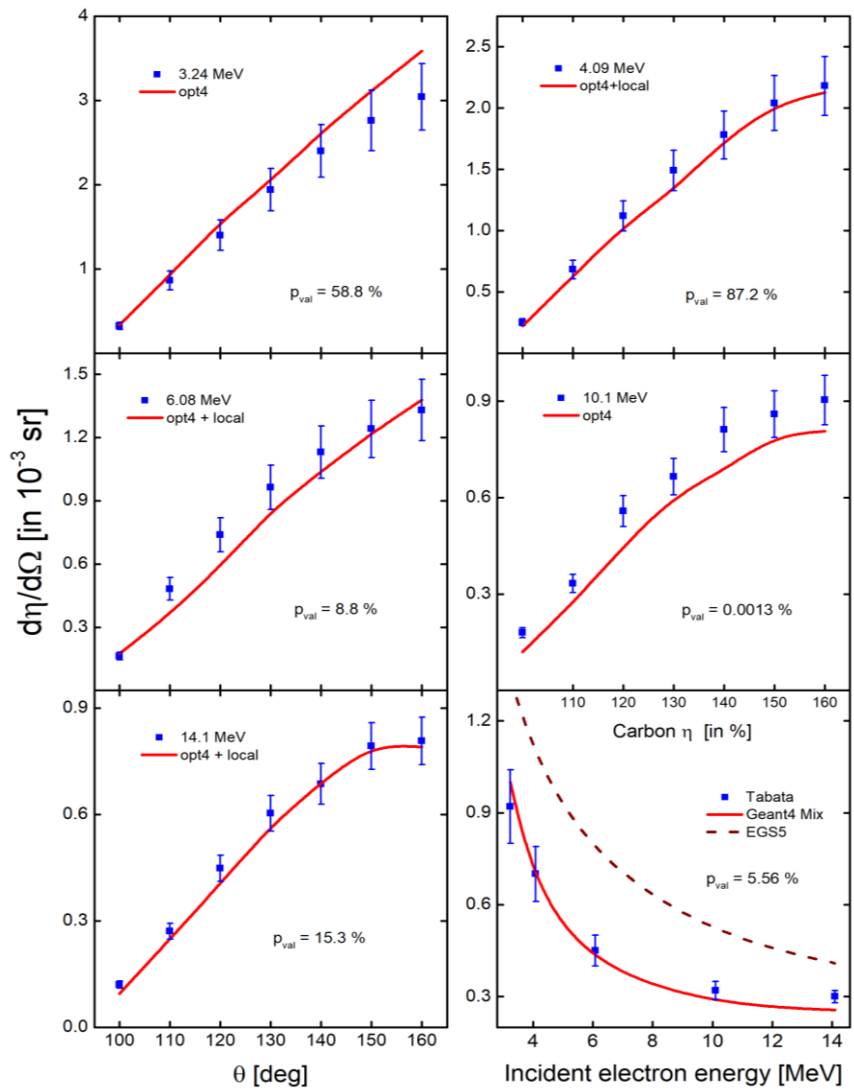

Fig. 3. Angular distributions and total backscattering coefficient in Carbon.

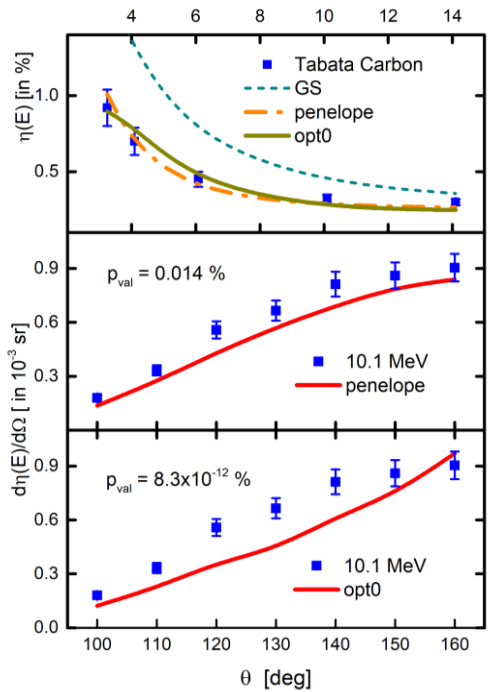

Fig. 4. Bad performance of some PLs in Carbon.

### D. Aluminium Results

It is clear from Fig. 5 that in Aluminium (Z=13) the discrepancy with the EGS5 calculation is smaller. The rejection of the calculations at 6.08 MeV is a bit puzzling in spite of the difference at 160°. This is a problem with the linear scale which masks the big contributions to the $\chi^2$ statistic as can be seen in Fig. 6.

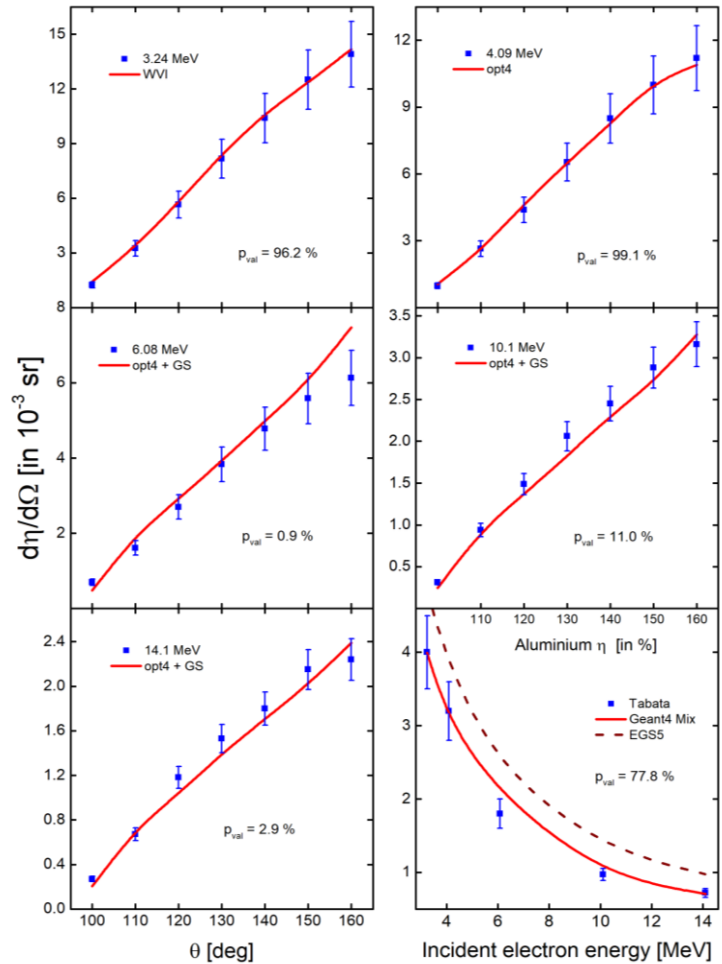

Fig. 5. Angular distributions and backscattering coefficients in Aluminium.

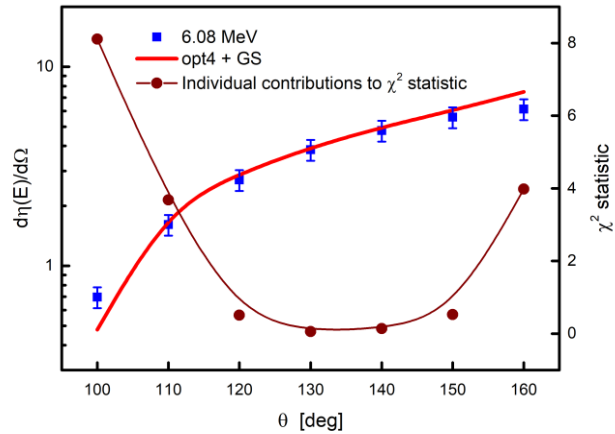

Fig. 6. Angular distribution at 6.08 MeV in Aluminium.

There is a big contribution of ≈8 to the statistic at 100°, even greater than the one at 160° of 4. In fact, it is the only point calculated with opt4 because the GS was even worse. The logarithmic scale masks a big difference at 110°, 1.610 measured vs. 1.950 calculated. The other 4 points give a negligible contribution. The GOF is a very tough test.

### E. Gold Results

The work of Tabata included also results for heavier nuclei. Not discrepancies in regard to the EGS5 predictions were found, by Kirihara, for these nuclei. Gold is added here to present the agreement, in a heavy nucleus, with the EGS5 calculations, in Fig. 7. The agreement is quite good with a single rejection at 10.1 MeV at the level of 2%. The calculations were performed with Molière formalism for MSC, can be that GS will do better. But a big surprise is the quality of agreement obtained with a Geant4 calculation in Fig. 8, the p-values being higher than 75%.

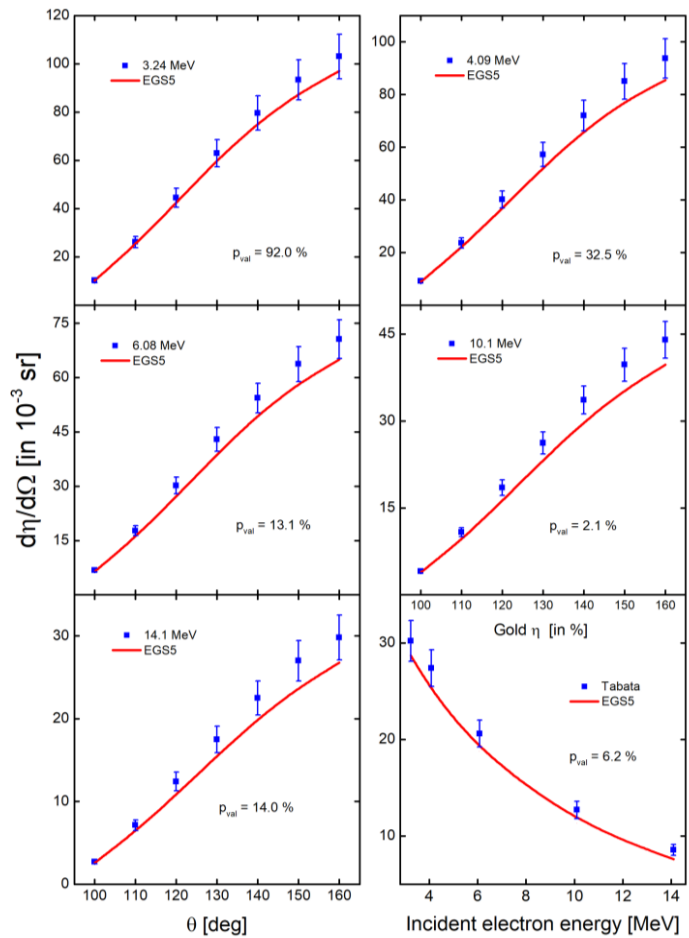

Fig. 7. Gold data compared with the EGS5 calculations.

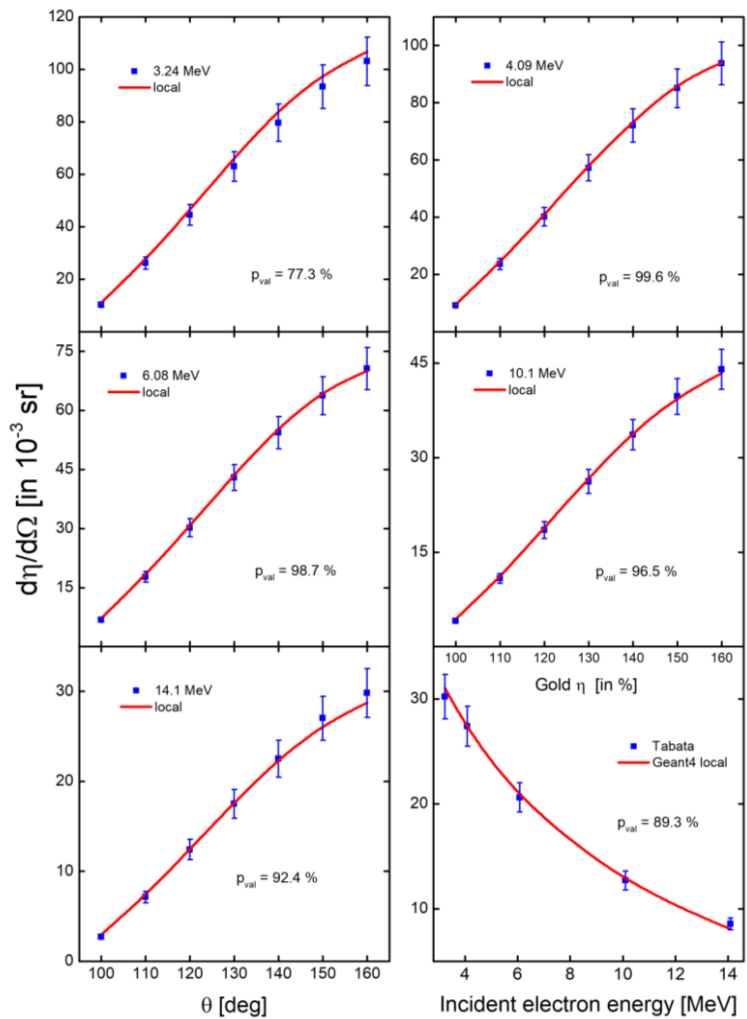

Fig. 8. Gold data compared with Geant4 + local PL calculations.

*F. EGS5 Program*

A few EGS5 calculations were presented here using a user program written on the basis of the KEK distribution with some modifications. All the COMMON statements were transformed in modules making the programs less cluttered. Four libraries were constructed: libegs5, libpegs5, libcg – for the combinatorial geometry and libaux – for some auxiliary routines not always needed. In the user part of the program allocatable vectors and derived types (TYPEDEF) were employed, taking advantage of programming directly in FORTRAN95, without the need of the cumbersome MORTRAN interpreter.

### III. Discussion

Kim *et al.* [**Error! Bookmark not defined.**] used a comprehensive approach in comparing a large number of experimental data with calculations based on Geant4. In this work I used a very fine resolution approach looking at a particular experiment. Nevertheless, it is of interest to compare the respective results. From the three low Z elements, addressed in the present work, only Carbon is also present in Kim's work, particularly in their Fig. 4. This is a complicated figure containing much experimental data and numerical calculations. Above ~1.5 MeV the only experimental points are those of Tabata's and one point from Heinrich [21]. The figure contains three calculations, performed with Geant4.9.1p03, noted as URBAN, URBANB and URBANBRF depending on the MSC model used. All of them are using the default Geant4 MSC model of Urban [**Error! Bookmark not defined.**] but differ in the method of limiting the step length. This region of energies is very compressed in Kim's figure due to the inclusion of a large number of experimental points at lower energies. The figure y scale goes up to values of η higher than 20% while the values in the MeV region of interest are less than 0.5%. Moreover, unfortunately, Kim *et al.* [1] are using quite large symbols both for experimental points and for the numerical calculations, making hard to really distinguish and digitize their values. This situation is the same, or worse, also in other figures, appearing Kim's paper and dealing with Carbon, namely Fig. 13, Fig. 27, Fig. 32 and Fig. 37, in which it is impossible to digitize their values at all. These additional figures differ by the version of Geant4 used for the calculations or/and by the MSC model.

In Fig. 4 of Kim, there is not a real difference between URBAN and URBANB, up to the size of the symbols representing them. Our Fig. 9 presents the results of Kim in the region of energies relevant to Tabata's experiment.

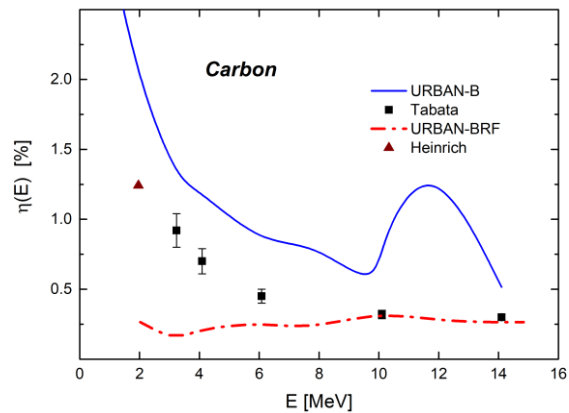

Fig. 9. Carbon data and calculations results from Kim *et. al*. [1].

The great bump at 11-12 MeV, seen in the URBANB curve above, it is a real feature present in Kim's calculations, not clear why. While the curve URBANBRF gives a nice agreement at 10.1 and 14.1 MeV, from a full look at Kim's figure, in its full range of energies, it is clear that it is the worst option for describing the MSC. Overall, the URBANB curve gives a poorer fit to Tabata's data compared with the fit obtained in Fig. 3.

Dondero *et al*. [22] also published Geant4 simulations of the electron backscattering including a number of different options for the MSC. They include only calculations <= 1 MeV and no comparison with the present work can be done.

The present paper presents solid evidence that the claim made by Kirihara [**Error! Bookmark not defined.**] et al., namely that there is a normalization problem in the electron backscattering measurements of Tabata [**Error! Bookmark not defined.**], for the low Z targets, it is not supported by calculations made with Geant4, up to carefully choosing the MSC model employed. EGS5 does not offer the wide range of options in choosing the MSC model present in Geant4. Both the successes or the failures of Geant4 and EGS5 calculations in reproducing the experimental data imply that the problem of accurately describing the electron MSC scattering does not yet have a solid solution in the Mote Carlo simulations. Moreover, I don't think that is right affirming failures in an experiment on the basis of not fitting a particular theory. It will be worthwhile expanding the MSC possibilities in EGS5 because, from the experience of this author, it is much more rapid compared with Geant4.

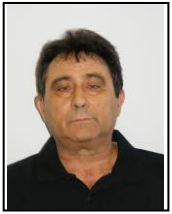
**Sylvian Kahane**. Born 1946 Bucharest, Romania. BSc (1968) and MSc (1971) Physics Technion - Israel Institute of Technology, Haifa, Israel. PhD (1977) Physics, Hebrew University, Jerusalem, Israel.
He retired from Physics Dept., Nuclear Rsearch Center – Negev in 2013. Interests: low energy (MeV) nuclear physics, atomic processes (Rayleigh scattering), gamma interactions, simulations, molecular dynamics, hyperfine interactions.